\documentclass[conference]{IEEEtran}
\IEEEoverridecommandlockouts
\usepackage{cite}
\usepackage{amsmath,amssymb,amsfonts}
\usepackage{algorithmic}
\usepackage{graphicx}
\usepackage{amssymb}
\usepackage{textcomp}
\usepackage{hyperref}
\usepackage{xcolor}

\def\BibTeX{{\rm B\kern-.05em{\sc i\kern-.025em b}\kern-.08em
    T\kern-.1667em\lower.7ex\hbox{E}\kern-.125emX}}
\begin{document}

\title{CSI-based Positioning in Massive MIMO systems using Convolutional Neural Networks}

\author{\IEEEauthorblockN{Sibren De Bast, Andrea P. Guevara, Sofie Pollin}
\IEEEauthorblockA{\textit{ESAT - TELEMIC} \\
\textit{KU Leuven}, Belgium\\
sibren.debast@kuleuven.be}
}

\maketitle

\begin{abstract}
This paper studies the performance of a user positioning system using Channel State Information (CSI) of a Massive MIMO (MaMIMO) system. To infer the position of the user from the CSI, a Convolutional Neural Network is designed and evaluated through a novel dataset. This dataset contains indoor MaMIMO CSI measurements using three different antenna topologies, covering a 2.5~m by 2.5~m indoor area. We show that we can train a Convolutional Neural Network (CNN) model to estimate the position of a user inside this area with a mean error of less than half a wavelength.
Moreover, once the model is trained on a given scenario and antenna topology, Transfer Learning is used to repurpose the acquired knowledge towards another scenario with significantly different antenna topology and configuration. Our results show that it is possible to further train the CNN using only a small amount of extra labelled samples for the new topology. This transfer learning approach is able to reach accurate results, paving the road to a practical CSI-based positioning system powered by CNNs.
\end{abstract}

\begin{IEEEkeywords}
Massive MIMO, Positioning, CSI, Deep Learning, Transfer Learning
\end{IEEEkeywords}

\section{Introduction}


Massive MIMO (MaMIMO) is a emerging technology used in 5G communication networks to greatly enhance the spectral efficiency of our wireless systems \cite{mamimo}. It does so by combining a large number of Base Station (BS) antennas with signal processing based on measured Channel State Information (CSI). This CSI is estimated using pilot sequences during up-link transmission. The combination of a large number of antennas and accurate CSI allows the BS to multiplex the users in the spatial domain. Theory shows that, as the number of base station antennas increases, the performance of the system is only bounded by the accuracy of the channel state information. If such accurate channel state information is available, the question arises if that information can be used to infer other context information about the environment.

CSI contains spatial information which is used by the BS to multiplex users in the spatial domain. Consequently, it can be extracted and used to localise the users in space. Localisation of wireless terminals has many interesting applications for both indoor as outdoor networks. For example, indoor navigation systems can be used to navigate users through buildings, and autonomous driving is enabled in covered areas where no GPS signals can be detected\cite{5gpos}. Furthermore, the CSI is already needed by the MaMIMO system in order to communicate, so there is no extra cost in using the CSI to localise the users.


Savic and Larsson introduced the notice for fingerprint-based position services in MaMIMO systems \cite{savic}. They propose several positioning methods based on classical machine learning algorithms ($\kappa~\text{Nearest Neighbours}$, Support Vector Machines, and Gaussian Process Regression). As input to these learning algorithms a vector containing the received signal strength values is used. However, this approach uses only a small part of the information available in a MaMIMO system.





In a paper by Vieira et al. \cite{lund}, the ability of using Convolutional Neural Networks (CNN) to extract the spatial information in the CSI of MaMIMO systems was studied. The CSI gives access to more information in comparison to a received signal strength vector, while the CNN provides an efficient way to process this data. They found that their model could reach below wavelength accuracy on a simulated test set of MaMIMO CSI samples. They evaluate their method with CSI samples generated by the COST 2100 MIMO model \cite{cost2100}, giving them access to an infinite amount of perfectly labelled data. With this set-up, the authors reach a performance of around 0.6 wavelengths. The real challenge, however, lies in gathering real-life labelled data. 

In \cite{sounder}, Arnold et al. present a novel channel sounder architecture to address this lack of measured MaMIMO CSI. They apply the dataset acquired by the proposed channel sounder with 64 antennas on the problem of indoor localisation. However, their results only reach an accuracy of around 75 cm, which is significantly less accurate than the proposed accuracy of Vieira \cite{lund}. This lower accuracy can be attributed to the low accuracy of the provided labels, which is suggested to be around 10 cm. 

Relaxing the need for highly-accurate labelled data will be the key for a practical deployment of a CNN-based localisation system. In current studies, each new scenario needs a vast amount of labelled data to train the CNN. The cost of gathering these amounts of data is too high to make fingerprint-based positioning techniques viable. For example, for a fully distributed MaMIMO architecture, the topology of the antennas will differ in function of the location where it is deployed. If for each new deployment a very large amount of labelled data is needed, this method will never reach a practical state.

This work will focus on gathering highly-accurate labelled measured data and minimising the need for this data when a new scenario is encountered. The main contributions of this paper are:

\begin{itemize}
  \item The creation of a highly-accurate spatially labelled CSI dataset of a state-of-the-art MaMIMO system which has been made publicly available,
  \item The application of CNNs to infer the location of the user, based on the measured CSI dataset,
  \item The transfer of knowledge between two scenarios with a different antenna topology to minimise the need for new labelled samples,
  \item State-of-the-art localisation abilities using the CSI of a MaMIMO system, reaching a mean accuracy of 23.92 mm, corresponing to 0.209 $\lambda$. This is an error 65\% lower than the accuracy reported by Vieira \cite{lund}.
\end{itemize}

%
%

\section{Dataset}

For the purpose of this study, a novel dataset was created. The dataset contains the CSI measured by the KU Leuven Massive MIMO testbed for many user positions. The Base Station (BS) is equipped with 64 antennas, which can transmit or receive simultaneously. These 64 antennas are used to receive a predefined pilot signal from each user. The CSI is estimated based on these pilot signals. The pilot tone consists of 100 subcarriers, evenly spaced in frequency. Therefore, the measured CSI can be represented by the matrix $\mathbf{H} \in \mathbb{C}^{N \times K}$ with N being the number of antennas at the BS and K the number of subcarriers. For more details about the system, the National Instruments Massive MIMO Application Framework documentation \cite{NI} can be consulted.

During the measurements, four single-antenna users were positioned indoor in an office. They were moved by CNC XY-tables \cite{openbuilds} along a predefined route. By using these XY-tables the error on the positional label is less than 1 mm. This route went along a grid spanning a 1.25 m by 1.25 m area for each user. All routes were fully in Line-of-Sight (LoS). Using this set-up, the area was scanned with 5 mm intervals, resulting in a dataset containing 252004 CSI samples with location accuracy of less than 1 mm.

Furthermore, the testbed's BS is designed to be very flexible in the deployment of the antenna array. This allowed for the creation of three different datasets, each with a unique antenna deployment. First, a Uniform Rectangular Array (URA) of 8 by 8 antennas was deployed. Second, Uniform Linear Array (ULA) of 64 antennas on one line was deployed. Finally, the antennas were distributed over the room in pairs of eight, making up the distributed scenario. To the best of our knowledge, the resulting dataset is the largest indoor MaMIMO CSI dataset with spatial labels.

\begin{figure}[!ht]
    \centering
    \includegraphics[width=\linewidth]{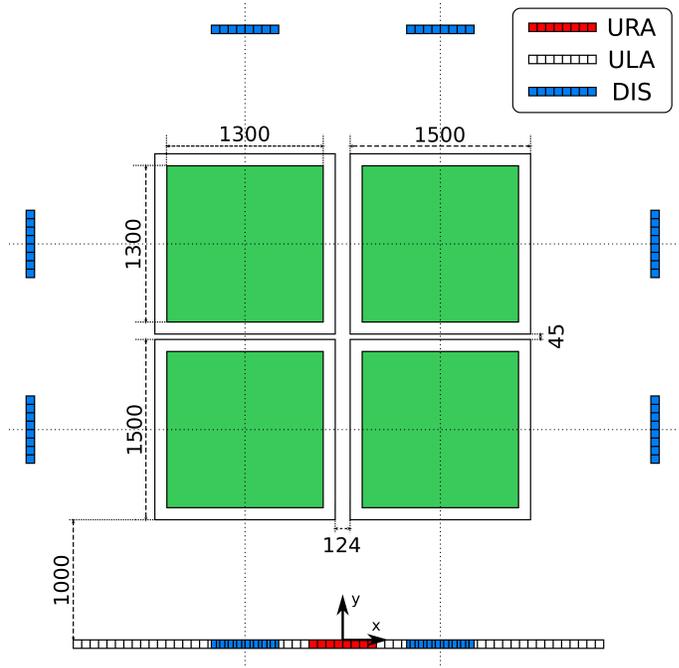}
    \caption{The three measurement scenarios. The antennas are spaced around the users using a scenario specific topology. The users are positioned inside the green areas. All measurements on the Figure are in $mm$.}
    \label{fig:scen}
\end{figure}

The different deployments can be seen in Figure \ref{fig:scen}. In all cases, the antennas are placed 1 m from the XY-tables. The green rectangles on the figure depict the 1.3 m by 1.3 m areas where the XY-tables are able to move the users in. To mitigate errors during the measurements, the centre 1.25 m by 1.25 m areas are used. The spacing in between the XY-tables is dictated by the space needed for the motors powering the movements and the cables connecting them to the controllers. These tables were synchronised over Ethernet with the BS to ensure the sampled $\mathbf{H}$ has a correct spatial label, enabling a highly accurate dataset.

\begin{figure}[!ht]
    \centering
    \includegraphics[width=\linewidth]{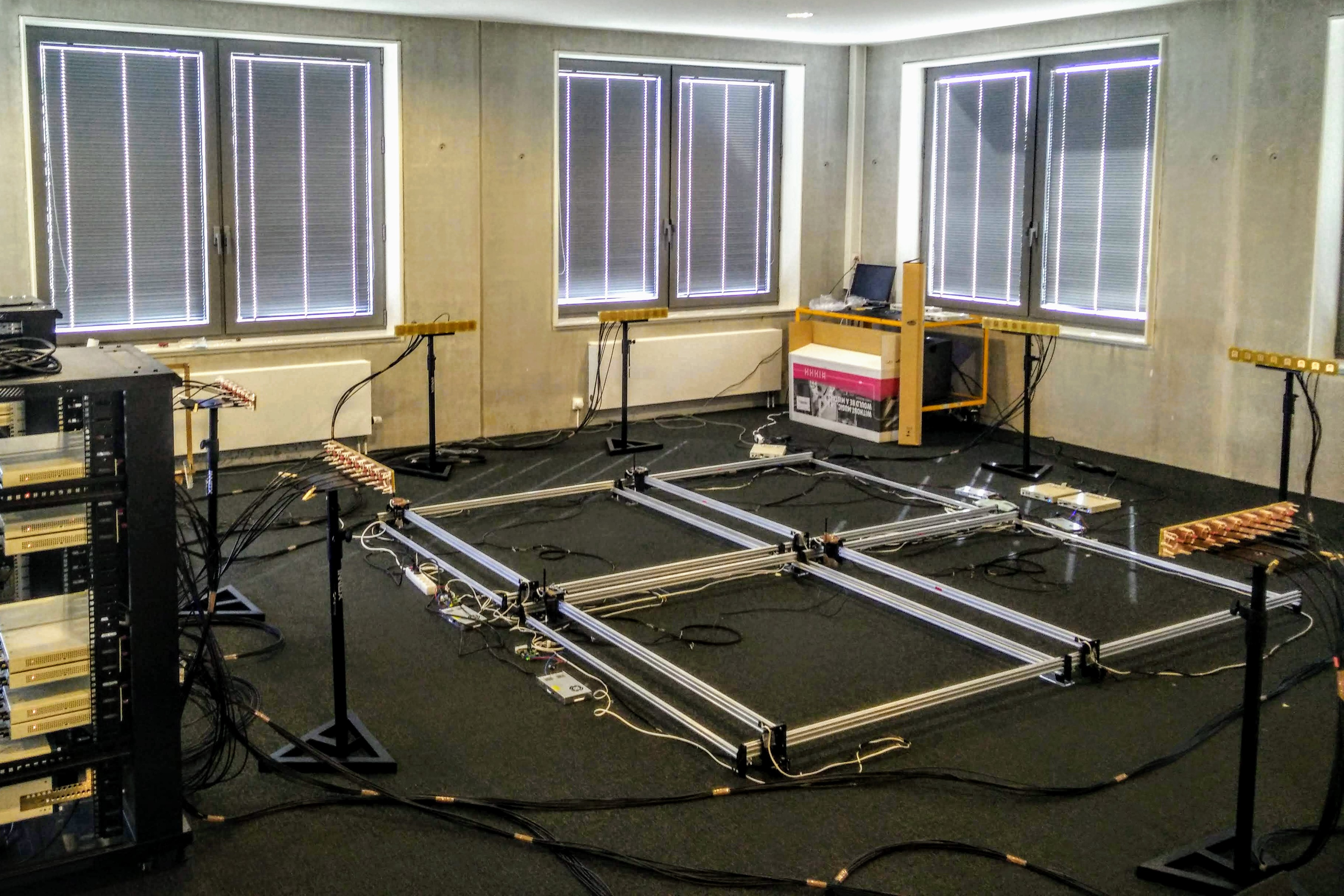}
    \caption{The measurement environment during the distributed Massive MIMO experiments. In the middle, the four XY-tables are shown, surrounded by the eight distributed antenna arrays.}
    \label{fig:photo}
\end{figure}

During the measurements, the BS was configured to use as centre frequency $f_c$ of 2.61 GHz, giving a wavelength $\lambda$ of 114.56 mm. The system used a bandwidth of 20 MHz. The spacing between adjacent antenna elements was 70 mm and the lowest antenna elements were located 93 cm above the floor. The user's antenna was placed 20 cm above the floor. The origin of the space was defined as the middle of the URA. From this point in space, the x- and y-positions of the users were measured. A picture of the distributed scenario can be seen on Figure \ref{fig:photo}.

Publicly available datasets containing spatially labelled MaMIMO CSI samples are very valuable to test and benchmark different MaMIMO CSI-based positioning methods. Furthermore, these datasets can be used for many more applications and studies exceeding the localisation problem. However, the amount of these kind of datasets is very limited. Therefore, to encourage further research, this dataset is made publicly available\footnote{\url{https://homes.esat.kuleuven.be/\~sdebast/}}.

\section{Convolutional Neural Network Model}

Convolutional Neural Networks (CNNs) have lately revolutionised the image classification field. They prove to be very efficient in learning relevant features in structured data to classify the data's content. The obtained MaMIMO CSI contains large amounts of structured data, as a result, CNNs make a good candidate technology to process these CSI samples and infer their spatial information. This section handles how the CNN for this specific task was designed. First, we evaluate how domain specific knowledge can be exploited to help the CNN extract useful features. Second, the architecture of the model is discussed and the Python code for the CNN is open-sourced.



\subsection{Domain specific knowledge}

While implementing the CNN for this task, domain specific knowledge can be used to help the network learn useful features. First of all, the dataset contains complex valued numbers. To help the CNN access this information, these complex valued numbers were converted to the polar domain. In this domain, the amplitude and the phase shift of the different subcarriers can easily be extracted, helping the CNN learn useful features. Second, since the information is presented in the frequency domain, we perform an Inverse Fast Fourier Transform on the data to reveal features in the time domain. Afterwards, these three different sets of features (raw features, polar features, and time-domain features) are concatenated before they are presented at the input of the CNN. This preprocessing results in an input matrix $\mathbf{I} \in \mathbb{R}^{[N \times K \times 6]}$.

When designing CNNs, the size of the convolutions can be chosen freely, this size is called the kernel size. When training a CNN for an image classification task, this size is often chosen as $(3, 3)$ or something similar. However, since this data is very different from image data and domain knowledge is available, the kernel sizes should be adjusted accordingly. Therefore, in the higher layers of our CNN, the kernel sizes are designed to perform a 1D convolution over the data. This way, the neural network first extracts relevant features out of the data for each antenna. The lower layers perform convolutions in the other dimensions to combine the features from multiple antennas. Designing the kernel sizes with domain specific knowledge can reduce the number of trainable parameters, which makes the CNN faster to train and less prone to overfitting.

\subsection{Architecture}

The full CNN contains 13 convolutional layers, improved with skip connections \cite{resnet} and drop-out layers, and three fully connected layers at the end. The total amount of trainable parameters depends on the number of antennas used in the model. For 64 antennas, the number of trainable parameters is $217,378$, which is only 1.36\% of the weights used by Arnold et al. \cite{sounder}.


Previous work on CSI-based positioning using CNNs did not include a detailed description of the applied Deep Learning techniques. Therefore, it is impossible to compare the learning efficiency of our method to others. To mitigate this problem in the future, and for more details on the implementation, the code has been published on-line. The CNN was implemented using Keras and TensorFlow and the source code for this research can be found at GitHub\footnote{\url{https://github.com/sibrendebast/MaMIMO_CSI_with_CNN_positioning}}. 


\section{Performance Evaluation}

This section explores the various factors influencing the positioning-ability of the aforementioned CNN. First, the performance is evaluated for training the CNN on the positioning task in the three different scenarios. Second, the influence of the number of antennas used at the BS is studied. Afterwards, the ability to use transfer learning to migrate knowledge from one scenario to another is explored. All results are computed on the testset, which is 5\% of the total dataset.

\subsection{Performance differences due to antenna topology}

\begin{figure}[!ht]
    \centering
    \includegraphics[width=\linewidth]{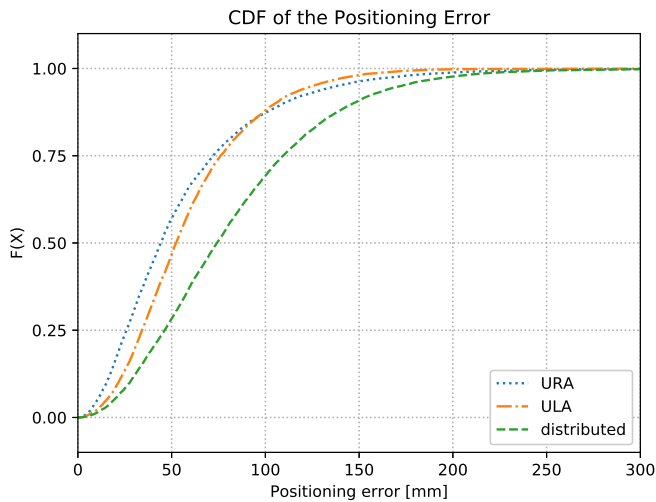}
    \caption{A CDF of positioning error on the three different scenarios}
    \label{fig:topology}
\end{figure}

The used antenna topology influences the accuracy of the positioning system. Therefore, three different antenna topologies are evaluated. Figure \ref{fig:topology} shows the CDF of positioning error of the three different antenna topologies. The model was trained on 85\% of the datasets, leaving 10\% for validation. The results show that the CNN is able to infer the spatial information for all three scenarios, with the URA and ULA scenario having the highest accuracy.

In the used dataset, the user are all placed at the same height above the floor. Therefore, the model does not need to infer the height of the users. However, the ability to infer the height information might be dependent on the antenna topology. Therefore, to further investigate the topology-dependence of the positioning performance, a dataset containing samples with varying height of the user is needed.

\subsection{Influence of the number of Antennas}

The number of antennas used in the BS also largely influences the accuracy. The authors of \cite{crlb} suggest that in a CSI-based positioning system, the accuracy of the system will rise with the amount of used antennas. This is exactly what we see when we evaluate the proposed model with subsets of the dataset. Here, we subsample the dataset so we only use the data provided by a specific amount of antennas. When the number of antennas is increased from 8 to 64, the accuracy of the system improves 58\% to 76\%, depending on the topology.

Table \ref{tab:accuracy} shows the Mean Error (ME) on the testset in function of the number of used antennas for the three scenarios. The ME is calculated as following:
$$\text{ME} = \mathbb{E}\{|p - \hat{p}|\},$$
where $p$ is the measured position $(x,y)$ of the user and $\hat{p}$ the estimated position.
The results are both shown in absolute accuracy, using millimetre as unit, and in relative accuracy, using one wavelength $\lambda$ as the unit. This allows to compare the results independently from the used carrier frequency.

\begin{table}[!ht]
    \centering
    \caption{Mean error the positioning performance of the proposed system, measured on the testset of the different scenarios.}

    \begin{tabular}{|l||r|r|r|r|}
    \hline
    URA & 8 antennas & 16 antennas & 32 antennas & 64 antennas \\
    \hline
    ME [mm] & 230.08 & 120.73 & 85.19 & 55.35 \\
    ME [$\lambda$] & 2.008 & 1.053 & 0.744 & 0.483 \\
    \hline
    \end{tabular}

    \vspace{3mm}

    \begin{tabular}{|l||r|r|r|r|}
    \hline
    ULA & 8 antennas & 16 antennas & 32 antennas & 64 antennas \\
    \hline
    ME [mm] & 244.80 & 132.87 & 77.87 & 59.05 \\
    ME [$\lambda$] & 2.137 & 1.160 & 0.680 & 0.515 \\
    \hline
    \end{tabular}

    \vspace{3mm}

    \begin{tabular}{|l||r|r|r|r|}
    \hline
    DIS & 8 antennas & 16 antennas & 32 antennas & 64 antennas \\
    \hline
    ME [mm] & 197.62 & 152.37 & 130.64 & 82.30 \\
    ME [$\lambda$] & 1.725 & 1.33 & 1.14 & 0.718 \\
    \hline
    \end{tabular}

    \label{tab:accuracy}
\end{table}

The results show that the accuracy of the system indeed improves with the amount of antennas used. When using 64 antennas an accuracy of around $\frac{1}{2}\lambda$ is reached for the URA and ULA topology. Now, the main question is to what extent the learned model can generalise over different contexts, such as different antenna topologies.

\subsection{Transfer Learning}

Using Transfer Learning \cite{transfer}, knowledge gathered during training for a previous task can be used to accelerate training on a similar new task. Furthermore, the number of necessary training samples to train the model for the new task is greatly reduced. This technique is often used for image classification, where the weights of the first layers in the CNN are reused in a new network. These first layers contain filters for low level features and they are quite generally applicable on similar tasks. In this way, the CNN already starts with a basic knowledge.

In this paper, the ability to transfer knowledge between two scenarios with a different antenna topology is studied. Specifically, the knowledge transfer for the URA scenario to the ULA scenario with 64 antennas is evaluated. The main idea is that the first layers of the CNN contain knowledge how to infere spatial relevant features from the raw data. This can then be reused, no matter which 
antenna topology is used. The lower layers than combine the separate spatial features to estimate the position of the user. Therefore, only the lower layers need to be retrained, which enables a faster training with a lower need for samples.

\begin{figure}[!ht]
  \centering
  \includegraphics[width=\linewidth]{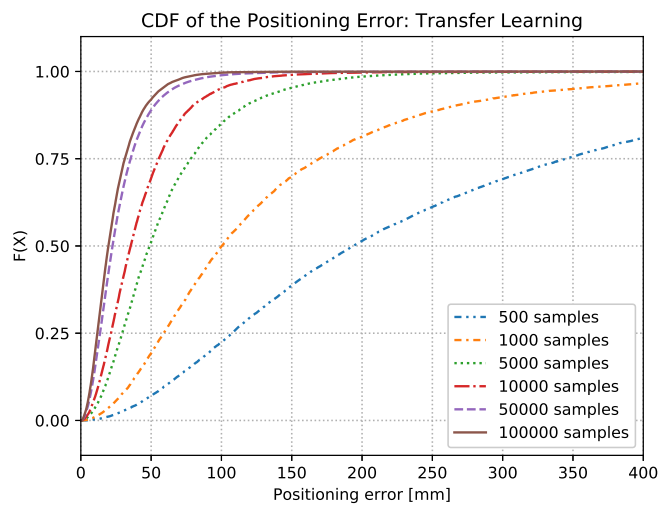}
  \caption{CDF of the mean error using a varying amount of training samples to perform transfer learning.}
  \label{fig:tl}
\end{figure}

Figure \ref{fig:tl} shows how we can achieve similar or even higher performance with a lower amount of training samples.
With only 5000 samples of the ULA training set, the CNN achieves a performance similar to the network trained on the full ULA training set without using transfer learning. Moreover, when 100000 samples are used for training, the system reaches a higher performance than the case when all ULA samples were used, but no transfer learning was applied. This is explained by the model having access to more samples to learn from than the case without transfer learning. This allows the model to further fine-tune the general statistics of the samples, while also learning the mechanics of the new scenario.

\begin{table}[!ht]
  \centering
  \caption{Mean error when using transfer learning in function of the number of used training samples.}
  \begin{tabular}{|l||r|r|r|r|r|r|}
  \hline
  \# samples & 500 & 1000 & 5000  & 10000 & 50000 & 100000 \\
  \hline
  ME [mm] & 253.5 & 132.0 & 60.90 & 42.79 & 27.89 & 23.92 \\
  ME [$\lambda$] & 2.21 & 1.15 & 0.53 & 0.37 & 0.243 & 0.209 \\
  \hline
  \end{tabular}
  \label{tab:tl}
\end{table}

Table \ref{tab:tl} shows the mean positioning error when using transfer learning from the URA to the ULA scenario. It demonstrates that transfer learning can reach a higher positioning performance than the case where only information of one specific scenario is used. 
The proposed technique reaches a ME of 23.92 mm, which corresponds to 0.209 $\lambda$.

\begin{figure}
    \centering
    \includegraphics[width=0.95\linewidth]{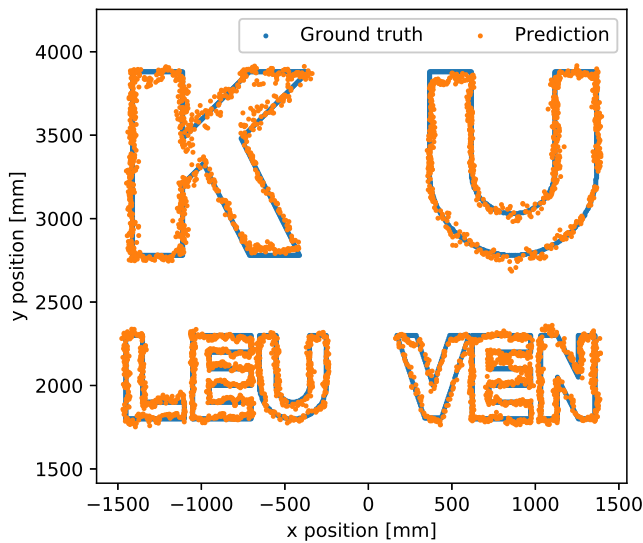}
    \caption{A visualisation example to show the performance of our proposed model. The model was first trained on the full URA training set, afterwards, using transfer learning, it was retrained with 100000 samples of the ULA training set. The visualisation was created using the ULA dataset.}
    \label{fig:leuven}
\end{figure}

To visualise the performance of our proposed model, the letters our university's name were converted to coordinates inside the areas of the XY-tables. The measured CSI at these locations was was used by the model to predict the corresponding locations. The model's predictions spell our univesity's name ``KU Leuven'' with a ME of 23.97 mm (0.21 $\lambda$) and are shown in Figure \ref{fig:leuven}. 

\section{Future work}

The first next step towards a practical implementation of the proposed techniques would be to evaluate different environments. Up until now, the scenario was changed by deploying the antennas in a different topology. However, the impact of transferring the system to another environment, e.g. another room, has to be studied as well. This will give extra insight into how the system infers the spatial information and how much data has to be gathered in a new environment to retrain the CNN using transfer learning.

Furthermore, recent advancements in semi-supervised learning allow for training a model with data that is only partially labelled. Since acquiring spatially-labelled data in new environments is very costly while gathering unlabelled data is very easy, reducing the need to label the gathered data, will reduce the deployment cost of the proposed positioning system. The main idea behind semi-supervised learning is that the model learns the statistics of the task using the unlabelled data, while the labelled data takes care of mapping this gathered knowledge to a useful output. Therefore, applying semi-supervised learning on this task can really push this techniques from research to a practical solution.

\section{Conclusion}

We investigated the ability of convolutional neural networks to infer the position of a user in a Massive MIMO communication system. The spatial information of the user was extracted from the channel state information gathered by the Base Station. To train the CNN a novel dataset of measured indoor MaMIMO CSI was created and published. To our knowledge, this dataset is the largest publicly available dataset with spatial labels, it consists of three different scenarios with each 252,004 CSI samples.

This dataset, together with a state-of-the-art CNN architecture, enabled us to infer the position of a user with an accuracy of 55.35 mm, which relates to 0.483 $\lambda$. This result outperforms any results found in literature. Furthermore, we studied the ability to use transfer learning to lower the amount data needed when the system is deployed with a different antenna topology. We show that the amount of labelled data needed to reach a similar performance, is reduces 20-fold in comparison to not using knowledge transfer. Furthermore, when more labelled data of the new scenario is available, this techniques can even reach a higher performance of 23.92~mm or 0.209 $\lambda$.

\section{Acknowledgement}
This research was partially funded by the Research Foundation Flanders (FWO) SB PhD fellowship, grant no. 1SA1619N (Sibren De Bast).

\end{document}